\documentstyle[prd,aps,floats]{revtex}

\begin{document}
\preprint{astro-ph/0002397}
\draft
\tighten

% Remove this and closure after abstract, plus preprint number,
% in electronic submission
%
\input epsf

\def\la{\mathrel{\mathpalette\fun <}}
\def\ga{\mathrelbe {\mathpalette\fun >}}
\def\fun#1#2{\lower3.6pt\vbox{\baselineskip0pt\lineskip.9pt
        \ialign{$\mathsurround=0pt#1\hfill##\hfil$\crcr#2\crcr\sim\crcr}}}

\renewcommand\({\left(}
\renewcommand\){\right)}
\renewcommand\[{\left[}
\renewcommand\]{\right]}

\newcommand\del{{\mbox {\boldmath $\nabla$}}}

\newcommand\eq[1]{Eq.~(\ref{#1})}
\newcommand\eqs[2]{Eqs.~(\ref{#1}) and (\ref{#2})}
\newcommand\eqss[3]{Eqs.~(\ref{#1}), (\ref{#2}) and (\ref{#3})}
\newcommand\eqsss[4]{Eqs.~(\ref{#1}), (\ref{#2}), (\ref{#3})
and (\ref{#4})}
\newcommand\eqssss[5]{Eqs.~(\ref{#1}), (\ref{#2}), (\ref{#3}),
(\ref{#4}) and (\ref{#5})}
\newcommand\eqst[2]{Eqs.~(\ref{#1})--(\ref{#2})}

\newcommand\pa{\partial}
\newcommand\pdif[2]{\frac{\pa #1}{\pa #2}}
                        
\newcommand\ee{\end{equation}}
\newcommand\be{\begin{equation}}
\newcommand\eea{\end{eqnarray}}
\newcommand\bea{\begin{eqnarray}}

%Fields and their VEVs
\def\so{S_1}
\def\st{S_3}
\def\stb{\overline{S}_3}
\def\se{S_8}
\def\see{S'_8}
\def\sf{S_{15}}
\def\sfp{S'_{15}}

\def\vo{|S_1|^2}
\def\vt{|S_3|^2}
\def\vtb{|\overline{S}_3|^2}
\def\ve{|S_8|^2}
\def\vee{|S'_8|^2}
\def\vf{|S_{15}|^2}
\def\vfp{|S'_{15}|^2}

%units
\newcommand\yr{\,\mbox{yr}}
\newcommand\sunit{\,\mbox{s}}
\newcommand\munit{\,\mbox{m}}
\newcommand\wunit{\,\mbox{W}}
\newcommand\Kunit{\,\mbox{K}}
\newcommand\muK{\,\mu\mbox{K}}

\newcommand\metres{\,\mbox{meters}}
\newcommand\mm{\,\mbox{mm}}
\newcommand\cm{\,\mbox{cm}}
\newcommand\km{\,\mbox{km}}
\newcommand\kg{\,\mbox{kg}}
\newcommand\TeV{\,\mbox{TeV}}
\newcommand\GeV{\,\mbox{GeV}}
\newcommand\MeV{\,\mbox{MeV}}
\newcommand\keV{\,\mbox{keV}}
\newcommand\eV{\,\mbox{eV}}
\newcommand\Mpc{\,\mbox{Mpc}}

%astronomical
\newcommand\msun{M_\odot}
\newcommand\mpl{M_{\rm P}}
\newcommand\MPl{M_{\rm P}}
\newcommand\Mpl{M_{\rm P}}
\newcommand\mpltil{\widetilde M_{\rm P}}
\newcommand\mf{M_{\rm f}}
\newcommand\mc{M_{\rm c}}
\newcommand\mgut{M_{\rm GUT}}
\newcommand\mstr{M_{\rm str}}
\newcommand\mpsis{|m_\chi^2|}
\newcommand\etapsi{\eta_\chi}
\newcommand\luv{\Lambda_{\rm UV}}
\newcommand\lf{\Lambda_{\rm f}}

\newcommand\lsim{\mathrel{\rlap{\lower4pt\hbox{\hskip1pt$\sim$}}
    \raise1pt\hbox{$<$}}}
\newcommand\gsim{\mathrel{\rlap{\lower4pt\hbox{\hskip1pt$\sim$}}
    \raise1pt\hbox{$>$}}}

\newcommand\diff{\mbox d}

\def\dbibitem#1{\bibitem{#1}\hspace{1cm}#1\hspace{1cm}}
\newcommand{\dlabel}[1]{\label{#1} \ \ \ \ \ \ \ \ #1\ \ \ \ \ \ \ \ }
\def\dcite#1{[#1]}

\def\dslash{\not{\hbox{\kern-2pt $\partial$}}}
\def\Dslash{\not{\hbox{\kern-4pt $D$}}}
\def\Oslash{\not{\hbox{\kern-4pt $O$}}}
\def\Qslash{\not{\hbox{\kern-4pt $Q$}}}
\def\pslash{\not{\hbox{\kern-2.3pt $p$}}}
\def\kslash{\not{\hbox{\kern-2.3pt $k$}}}
\def\qslash{\not{\hbox{\kern-2.3pt $q$}}}

 \newtoks\slashfraction
 \slashfraction={.13}
 \def\slash#1{\setbox0\hbox{$ #1 $}
 \setbox0\hbox to \the\slashfraction\wd0{\hss \box0}/\box0 }
 
% EXAMPLE OF HOW TO USE IT
% $\slash D$
% {\slashfraction={.075} $\slash{\cal A}$}
% $\slash B$
% $\slash a$
% {\slashfraction={.09} $\slash p$}
% $\slash q$

\def\ee{\end{equation}}
\def\be{\begin{equation}}
\def\smallfrac#1#2{\hbox{${\scriptstyle#1} \over {\scriptstyle#2}$}}
\def\fourth{{\scriptstyle{1 \over 4}}}
\def\half{{\scriptstyle{1\over 2}}}
\def\st{\scriptstyle}
\def\sst{\scriptscriptstyle}
\def\mco{\multicolumn}
\def\epp{\epsilon'}
\def\vep{\varepsilon}
\def\ra{\rightarrow}
\def\ppg{\pi^+\pi^-\gamma}
\def\vp{{\bf p}}
\def\ko{K^0}
\def\kb{\bar{K^0}}
\def\al{\alpha}
\def\ab{\bar{\alpha}}

\def\calm{{\cal M}}
\def\calp{{\cal P}}
\def\calr{{\cal R}}
\def\calpr{{\calp_\calr}}

\newcommand\bfa{{\bf a}}
\newcommand\bfb{{\bf b}}
\newcommand\bfc{{\bf c}}
\newcommand\bfd{{\bf d}}
\newcommand\bfe{{\bf e}}
\newcommand\bff{{\bf f}}
\newcommand\bfg{{\bf g}}
\newcommand\bfh{{\bf h}}
\newcommand\bfi{{\bf i}}
\newcommand\bfj{{\bf j}}
\newcommand\bfk{{\bf k}}
\newcommand\bfl{{\bf l}}
\newcommand\bfm{{\bf m}}
\newcommand\bfn{{\bf n}}
\newcommand\bfo{{\bf o}}
\newcommand\bfp{{\bf p}}
\newcommand\bfq{{\bf q}}
\newcommand\bfr{{\bf r}}
\newcommand\bfs{{\bf s}}
\newcommand\bft{{\bf t}}
\newcommand\bfu{{\bf u}}
\newcommand\bfv{{\bf v}}
\newcommand\bfw{{\bf w}}
\newcommand\bfx{{\bf x}}
\newcommand\bfy{{\bf y}}
\newcommand\bfz{{\bf z}}

\newcommand\sub[1]{_{\rm #1}}
\newcommand\su[1]{^{\rm #1}}

\newcommand\supk{^{(K) }}
\newcommand\supf{^{(f) }}
\newcommand\supw{^{(W) }}
\newcommand\Tr{{\rm Tr}\,}

\newcommand\msinf{M\sub{inf}}
\newcommand\phicob{\phi\sub{COBE}}
\newcommand\delmult{\Delta V_{\chi\widetilde\chi{\rm f}}}
\newcommand\mgrav{m_{3/2}(t)}
\newcommand\mgravsq{m_{3/2}(t)}
\newcommand\mgravvac{m_{3/2}}

\newcommand\cpeak{\sqrt{\widetilde C_{\rm peak}}}
\newcommand\cpeako{\sqrt{\widetilde C_{\rm peak}^{(0)}}}
\newcommand\omb{\Omega\sub b}
\newcommand\ncobe{N\sub{COBE}}
\newcommand\vev[1]{\langle{#1}\rangle}
%@twocolumnfalse\endcsname]

\title{Observational constraints on the spectral index\\ of the cosmological
curvature perturbation }
\author{$^1$ David H.~Lyth, $^2$ Laura Covi}
\address{$^1$Physics Department, Lancaster University
Lancaster LA1 4YB, Great Britain}
\address{$^2$DESY Theory Group, Notkestrasse 85, D-22603 Hamburg, Germany}
%\date{February 2000} 
\maketitle
\begin{abstract}
 We  evaluate the observational constraints on the  spectral index $n$,
in the context of the $\Lambda$CDM hypothesis which represents the
 simplest viable cosmology. We first take 
 $n$ to be  practically scale-independent. Ignoring reionization,
we find at  a nominal 2-$\sigma$ level
 $n\simeq 1.0 \pm 0.1$. If we make
 the more realisitic assumption that 
 reionization occurs when a  fraction $f\sim 10^{-5}$ to $1$
  of the matter has collapsed, 
the 2-$\sigma$  lower bound is unchanged while the 1-$\sigma$ bound rises
slightly. These constraints  are 
 compared  with the 
prediction of various inflation models. Then we
 investigate the two-parameter scale-dependent spectral
index, predicted by  running-mass inflation models, and find
 that present data allow significant scale-dependence
of $n$, which occurs in a physically reasonable
regime of parameter space.
\end{abstract}

\pacs{PACS numbers: 98.80.Cq \hfill astro-ph/0002397}

\section{Introduction}

It is generally supposed that structure in the Universe originates
from a primordial gaussian curvature perturbation,  generated
by slow-roll inflation. The  spectrum $\calpr(k)$ of the curvature
perturbation is the point
of contact between observation and models of inflation. It is   given
in terms of the inflaton potential $V(\phi)$ by\footnote
{As usual, $\mpl=2.4\times 10^{18}\GeV$ is the Planck mass, 
$a$ is the scale factor and $H=\dot a/a$ is the Hubble parameter, and
$k/a$ is the wavenumber. We assume the usual slow-roll conditions
$\mpl^2|V''/V|\ll 1$ and $\mpl^2(V'/V)^2\ll1$, leading to 
$3H\dot\phi\simeq -V'$.}
\be
\frac4{25}\calpr(k)
 = \frac1{75\pi^2\mpl^6}\frac{V^3}{V'^2} \,,
\label{delh}
\ee
where the 
 potential and its derivatives are
 evaluated at the epoch of horizon exit
$k=aH$. To work out the value of $\phi$ at this epoch one uses
the relation
\be
\ln(k\sub{end}/k)\equiv N(k)
=\mpl^{-2}\int^\phi_{\phi\sub{end}} (V/V') \diff\phi
\,,
\label{Nofv}
\ee
 where $N(k)$ is actually the number
of $e$-folds from horizon exit  to the  end of slow-roll inflation.
At the scale explored by the COBE measurement of the cosmic microwave
background (cmb) anisotropy, 
 $N(k\sub{COBE})$ depends on the expansion of the Universe after inflation
in the manner specified by \eq{Ncobe} below.

Given this prediction, the
  observed
 large-scale  normalization  $\calp_\calr^{1/2}\simeq 10^{-5}$ provides a 
strong 
constraint on models of inflation.
Taking that for granted, we are here interested in the scale-dependence of
the spectrum, defined by the, in general, scale-dependent spectral 
index $n$;
\be
n(k)-1\equiv \frac {\diff \ln \calpr}{ \diff \ln k}
\,.
\ee
According to most inflation models, $n$ has negligible variation on
cosmological scales so that $\calpr\propto k^{n-1}$, but
we shall also discuss an interesting class of models giving
a different scale-dependence.

{}From \eqs{delh}{Nofv},
\bea
n-1 &=&  2\mpl^2 (V''/V)-3\mpl^2 (V'/V)^2 
\,,
\label{nofv}
\eea
and in  almost all models of inflation, \eq{nofv} is well approximated by
\be
n-1=2\mpl^2(V''/V)
\label{nofvapprox}
\,.
\ee
We see that the  spectral index 
 measures the {\em shape} of the inflaton potential $V(\phi)$,
being independent of its overall normalization. For this reason, 
it is a powerful discriminator between models of inflation.

The observational constraints on 
the spectral index have been studied by many authors, but 
a new investigation is justified for  two reasons.
On the observational side, the  
 cosmological parameters are at last being pinned down, 
 as is  the height of the first peak in the spectrum the cmb 
anisotropy.
No study has yet been given which takes on board  these observational
developments, while at the same time taking on board the crucial
influence of the reionization epoch on the peak height.
 On the theory
side, it is known  that the spectral index may  be strongly
scale-dependent if  the inflaton has a gauge coupling, leading
to what are called running-mass models. The quite specific, two-parameter
 prediction for 
the scale dependence of the  spectral index in these models
 has not been compared with
presently available data.

\section{The observational constraints on the parameters of the
$\Lambda$CDM model}
Observations of various types 
 indicate that we live in a low density 
Universe, which is at least approximately flat 
\cite{likely,url,boom,ten,maxima}.
 In the interest of  simplicity
we therefore adopt the 
 $\Lambda$CDM model, defined by the
requirements that the Universe is   exactly flat, and that the 
non-baryonic dark matter is cold with negligible interaction.
Essentially  exact flatness is predicted by inflation, unless one invokes
a special kind of model, or special initial conditions.
Also, there is no clear 
 motivation  to modify the cold dark matter hypothesis.\footnote
{In particular, the rotation curves  of dwarf galaxies may be
compatible with cold dark matter \cite{dwarf}.}
We shall constrain the parameters of the $\Lambda$CDM model, including
the spectral index, by performing a least-squares fit to 
 key observational quantities.

\subsection{The parameters}
The $\Lambda$CDM model is defined by the spectrum
$\calpr(k)$  of the
primordial curvature perturbation, and 
 the four   parameters that are
needed to translate this spectrum into spectra for
the  matter density perturbation and the cmb anisotropy.
The   four parameters are the
 Hubble constant $h$ (in units of 
$100\km\sunit^{-1}\Mpc^{-1}$),
 the total matter density parameter $\Omega_0$,  the 
baryon density parameter $\omb$, and the 
reionization redshift $z\sub R$. As we shall describe, 
 $z\sub R$ is  estimated by assuming that reionization occurs when some
fixed fraction $f$ of the matter collapses.
Within the reasonable range $f\sim 10^{-4}$ to $1$, the main results
are insensitive to the precise value of $f$.

The spectrum is conveniently  specified by its value at a scale explored
by COBE, and the spectral index $n(k)$.
We  shall  consider the usual case of  a constant spectral index,
and the case of running mass models where $n(k)$  is given by
a two-parameter expression.  Since $\calpr(k\sub{COBE})$ is determined
very accurately by the COBE data (\eq{cobe1} below) we fix its value.
Excluding   $z\sub R$ and $\calpr(k_{COBE})$,
  the $\Lambda$CDM model is  specified by
four  parameters in the case of a constant spectral index, or by five
parameters in the case of running mass inflation models.

\subsection{The data}
To compare the $\Lambda$CDM model with observation, we 
take as our starting point a 
study performed a few years ago \cite{llvw}. 
We consider the same seven  observational quantities as in the earlier
work, since they still summarize most  of the relevant
data. Of these quantities, three are
 the cosmological  quantities
$h$, $\Omega_0$, $\Omega\sub B$, which  we are also taking as free
parameters. The crucial difference between the present situation and 
the earlier one is that observation is beginning to pin down
$h$ and $\Omega_0$. Judging by the spread of measurements,
the systematic error, while still important, is no longer completely
dominant compared with the random error. At least at some
crude level, it therefore makes sense to pretend that the errors are all
random, and to perform a least squares fit.
The adopted values and errors
are given in Table 1, and summarized below. In common with earlier
investigations, we take the errors
to be uncorrelated.

\paragraph{Hubble constant}
On the basis of observations that have
nothing to do with large scale structure
it seems very likely \cite{likely} that $h$ is in the range $0.5$ to $0.8$. 
We therefore
adopt, at notionally the 2-$\sigma$  level, the value
 $h=0.65\pm 0.15$, corresponding to $h=0.65\pm 0.075$
at the notional 1-$\sigma$ level. 

\paragraph{The matter density}
The case of  the total density parameter $\Omega_0$ is similar
to that of the Hubble parameter. On the basis of observations that 
have nothing to do with large scale structure,
 it seems very likely \cite{likely} that  $\Omega_0$
lies between $0.2$ and $0.5$, and 
 we adopt  at the notional 1-$\sigma$ level
the value $\Omega_0 = 0.35 \pm 0.075$. 

\paragraph{The baryon density}
As described for instance in \cite{osw,subir}, the baryon density
parameter $\Omega\sub b$
has two likely ranges. At the  1-$\sigma$  level, these are estimated
in \cite{osw} to be $\Omega\sub b h^2 =.019\pm .002 $  and 
$\Omega\sub b h^2 =.007\pm.0015$. 
We adopt  the high  $\Omega\sub b$ range, which
is generally regarded as the most likely, though our conclusions would
be much the same if we were to adopt the low range.

\paragraph{The rms density perturbation at $8h^{-1}\Mpc$}
Primarily through the abundance of rich galaxy clusters,
 a useful constraint on the primordial spectrum is provided by
the  rms density contrast, in a comoving sphere with present radius
 $R\sim 10h^{-1}\Mpc$, at redshift $z=0$ to a few.
 The constrained quantity is conventionally taken to be the present,
linearly evolved rms density contrast at $R=8h^{-1}\Mpc$,
 denoted by $\sigma_8$.
A recent estimate \cite{vl} based on low-redshift clusters
gives at 1-$\sigma$
\bea
\sigma_8 &=& \widetilde \sigma_8 \Omega_0^{-0.47} \\
\widetilde \sigma_8 &=& .560\pm .059
\,.
\eea
This constrains the primordial curvature perturbation on 
the scale $k\sim k_8\equiv (8h^{-1}\Mpc)^{-1}$.

\paragraph{The shape parameter}
The slope of the galaxy correlation function on  scales of order
$1h^{-1}$ to $100h^{-1}\Mpc$ 
 is  conveniently specified  by a shape parameter \cite{llvw}
$\widetilde \Gamma$,
defined by
\bea
\widetilde \Gamma &=&  \Gamma - 0.28(n_8^{-1}-1) \\
 \Gamma &=& 
 \Omega_0 h \exp(-\Omega\sub B -\Omega\sub B/\Omega_0)
\,.
\eea
(The quantity $\Gamma$ determines, to an excellent approximation, the
shape of the 
matter transfer function on scales $k^{-1}\sim 1$ to $100h^{-1}\Mpc$,
while the second term accounts for the scale dependence of the 
primordial spectrum.
For  definiteness, we evaluate $n$ at
$k=k_8$, in  the case that $n$ has significant scale
dependence.)
A fit reported in
 \cite{llvw} gives $\widetilde \Gamma = .23$ with a $15\%$ uncertainty
at 2-$\sigma$. A more recent fit with more data
 \cite{will} gives $\widetilde \Gamma=.20 $ to $.25$, depending on the
assumed velocity dispersion, but with $15\%$ statistical
uncertainty at the 1-$\sigma$ level.\footnote
{See Table 3 of \cite{will}; in the present
context one  should focus on the last three rows of the Table.}
We therefore adopt $\widetilde\Gamma=.23$, with $15\%$ uncertainty
at 1-$\sigma$.

\paragraph{The  COBE normalization of the spectrum}
To a good approximation, the
 spectrum $C_\ell$ of the cmb anisotropy  at large $\ell$
is sensitive to the
primordial spectrum on  the corresponding scale at the particle
horizon, 
\bea
k(\ell,\Omega_0) &=& \frac{\ell}{x\sub{hor}(\Omega_0)} \label{kofell}\\
x\sub{hor}&\equiv& 2H_0^{-1} \Omega_0^{-1/2} \( 1+0.084\ln\Omega_0 \)
\label{xhor}
\,.
\eea
The COBE measurements cover the range $2\leq \ell \lsim 30$, and 
they constrain $\calpr(k)$ on the corresponding scales.
Instead of $\calpr$, it is usual in this context to consider a quantity
$\delta_H$, which is of direct interest for studies of 
structure formation and is  defined by
\bea
\delta_H(k) &\equiv
&\frac25 \frac{ g(\Omega_0)}{\Omega_0} \calpr^{1/2}(k) \label{delhdef}\\
g(\Omega_0)&\equiv&\frac52 \Omega_0 \(\frac 1{70}+\frac{209\Omega_0}{140}
-\frac{\Omega_0^2}{140} + \Omega_0^{4/7} \)^{-1}
\,.
\eea
The factor $g/\Omega_0$, normalized to 1 at $\Omega_0=1$,
  represents  the $\Omega_0$-dependence of the 
present, linearly evolved,  density contrast after pulling out
the scale-dependent transfer function and $\calpr$. Equivalently, 
$a(\Omega)g(\Omega)$ 
is the time-dependence of the density contrast after matter domination.

 According to the
ordinary (as opposed to 'integrated')   Sachs-Wolfe  approximation
\be
 C_\ell=\frac{4 \pi}{25}
 \int^\infty_0 \frac{\diff k}k j_\ell^2\( k x\sub{hor} \)
\calpr(k)
\,.
\ee
In the regime $\ell\gg 1$, it
 satisfies  \eq{kofell}  because $j_\ell^2$  peaks when its argument
is equal to $\ell$.
In the $\Lambda$CDM model, the Sachs-Wolfe  approximation is quite
good in  COBE regime, but still the quality of the 
data justify using the  full (linear) calculation, given for instance by the
output of the CMBfast package \cite{CMBfast}.

Consider first the case
  $n=1$ (scale-independent spectrum).
In the Sachs-Wolfe approximation, the value of $\calpr$
obtained by fitting the  COBE  data  is  independent
of  the cosmological parameters $h$,
$\Omega_0$ and $\Omega\sub b$.
Using instead the full calculation, a fit to the data by 
  Bunn and White \cite{bw} gives
\bea
\delta_H &=&\Omega_0^{-0.785 -0.05 \ln\Omega_0} 
 \widetilde \delta_H \nonumber\\
10^5\widetilde \delta_H &=& 1.94\pm 0.08 \label{cobe1}
\,,
\eea
As expected, the 
 corresponding spectrum of the curvature perturbation
has only mild dependence on $\Omega_0$
($\calpr\propto \Omega_0^{-0.03}$).

Consider next the case of a scale-independent spectral index $n\neq 1$.
Dropping an insignificant term quadratic in $n-1$, the
 fit of Bunn and White \cite{bw} handles the $n$-dependence by assuming
that  \eq{cobe1}  holds at a   'pivot' scale $k\sub{COBE}$ which is
independent of $\Omega_0$.\footnote
{Keeping the quadratic term, the 'pivot' scale  at which \eq{cobe1} holds
is dependent on $n$, but still independent of $\Omega_0$.
A related fit by Bunn, Liddle and White \cite{blw} keeps a cross-term
in $(n-1)$ and $\Omega_0$, which makes  the 'pivot' scale increase
with $\Omega_0^{-1}$, though not as strongly as in \eq{kcobeofom}
below.}
\be
k\sub{COBE} \equiv 6.6H_0
\,,.
\label{kcobe}
\ee
Insofar as the approximation \eq{kofell} is valid, this corresponds
to fixing $C_\ell$ at an $\Omega_0$-dependent value of $\ell$, which is
 $\ell=13$ for $\Omega_0$, and   $\ell=22$ 
for our central value $\Omega_0=.35$.

In the case of a scale-independent $n$, an alternative fit is provided
by the CMBfast package, which chooses $\calpr(k)$ to fit an
$n$-independent best-fit value of $C_{10}$. As expected, the 
 output of CMBfast is in good agreement with the
 Bunn-White fit. Even better agreement is obtained
using 
\be
k\sub{COBE}(\Omega_0)\equiv 13.2/x\sub{hor}
\,,
\label{kcobeofom}
\ee
which reduces to \eq{kcobe} for $\Omega_0=1$. Insofar as \eq{kofell} is
valid, this $\Omega_0$-dependent pivot for $k$ corresponds to an 
$\Omega_0$-independent pivot for $\ell$, namely $\ell=13$.

We are also interested in the 
 scale-dependent $n$
 predicted by the  running-mass inflation models.
However, as 
 the range of scales  explored by COBE corresponds to only
$\Delta N\simeq 2$, with the central values of $\ell $ the most important,
we can take the variation of $n$ to be negligible on these scales.

Guided by these considerations, 
we have adopted three slightly different versions
of the COBE normalization, chosen for convenience according to the context.
 When calculating 
$\widetilde \Gamma$ and $\widetilde \sigma_8$, we
in all cases
fixed  $\delta_H$ at the central value  given by \eq{cobe1}, at
the Bunn-White pivot point  $k\sub{COBE}$.
When calculating the height of the first peak in the cmb anisotropy,
 in the case of 
 the running-mass model, we used
\eq{kofell}, with $\delta_H$ again fixed at the central
value given by \eq{cobe1} but now
evaluated  at the slightly more accurate
pivot point $k\sub{COBE}(\Omega_0)$.
Finally, when evaluating the peak  height in the case of scale-independent $n$,
we used a linear fit to the output of CMBfast.
Explicit expressions for the peak height will be given after
considering the effect of reionization.

\paragraph{The peak height}
The model under consideration predicts a peak in the cmb anisotropy at
$\ell\simeq 210$ to $230$, and 
presently available data \cite{url,boom,ten,maxima} confirm the 
existence of a peak at about this position.
We adopt as a crucial observational  quantity  
$\widetilde C\sub{peak}$, defined as the maximum value of 
\be
\widetilde C_\ell \equiv \ell(\ell +1) C_\ell/2\pi
\,.
\ee
Presently available data give conflicting estimates
\cite{url,boom,ten,maxima}
  of $\cpeak$, with central values in the range 70 to
$90\muK$.
 We  adopt
$(80\pm 10)\muK$ with the uncertainty  taken to be at 1-$\sigma$.

\subsection{Reionization}
The effect of reionization on the cmb anisotropy is determined by  the
optical depth $\tau$. 
 We assume  sudden, complete reionization
at redshift  $z\sub R$, so that the optical depth $\tau$ is given by
\cite{peacock,abook}
\be
\tau= 0.035\frac{\omb}{\Omega_0} h \( \sqrt{\Omega_0(1+z\sub R)^3 +1
-\Omega_0} -1 \)
\,.
\ee

In previous investigations,  $z\sub R$ has 
 been regarded as a free parameter, usually  fixed at zero or some
other value.  In this investigation, we instead take on board that fact that
 $z\sub R$ can be estimated, in terms of the parameters that we are varying
plus assumed astrophysics. Indeed,
it is usually supposed that reionization occurs at an early epoch,
when  some fraction
 $f$ of the matter has 
collapsed,  into objects with mass very roughly  $M=10^6\msun$.
Estimates of $f$ are in the range \cite{llreion}
\be
10^{-4.4}\lsim f\lsim 1
\label{fest}
\,.
\ee
In the case $f\ll 1$, the 
 Press-Schechter approximation gives the estimate 
\be
1+z\sub R \simeq\frac{\sqrt2 \sigma(M)}{\delta\sub c g(\Omega_0)}
{\,\rm erfc}^{-1}(f)
\hspace{4em}(f\ll 1)
\label{fll1}
\,.
\ee
Here  $\sigma(M)$ is the present, linearly evolved,
 rms density contrast  with   top-hat smoothing, 
 and $\delta\sub c=1.7$ is the overdensity required
for gravitational collapse. 
( $g$ is the suppression factor of the linearly
evolved density contrast at the present epoch, which does not apply
at the epoch of reionization.)
In the case $f\sim 1$,  one can justify only the 
rough estimate
\be
1+z\sub R \sim \frac{ \sigma(M)}{g(\Omega_0)}
\hspace{4em}(f\sim 1)
\,.
\ee
(This estimate is not very different from the one that would be obtained
by using $f=1$ in \eq{fll1}.)

In our fits, we fix $f$ at different values in the above range,
and find that the most important results are not very sensitive to $f$
even though the corresponding values of $z\sub R$ can be quite high.

\subsection{The predicted peak height}

The   CMBfast package \cite{CMBfast} gives $C_\ell$, 
for 
 given values of the parameters with $n$ taken to be 
scale-independent.
Following  \cite{martin}, we parameterize the 
CMBfast output at the first peak   in the form
\be
\cpeak = \cpeako \(\frac{220}{10}\)^{\nu/2}
\,,
\label{eq:cpeak}
\ee
where 
\be
\nu \equiv a_n(n-1)
 + a_h \ln(h/0.65) + a_0\ln(\Omega_0/0.35)
+a\sub b h^2(\omb - {\omb}^{(0)}) -0.65f(\tau)\tau
\,.
\ee
 $\cpeako$ is the value of $\cpeak$
evaluated with each term of $\nu$ equal to zero.
The 
coefficients for the high choice $\omb^{(0)}h^2=0.019$ are 
$a_n=0.88$, $a_h= -0.37$, $a_0=-0.16$,
$a\sub b= 5.4$, and $\cpeako=77.5\muK$. The formula
reproduces the CMBfast results within 10\% for a 1-$\sigma$
variation of the 
cosmological parameters, $h, \Omega_0$ and $ \omb$,
and $n\sub{COBE}=1.0\pm 0.05$.  
With the function  $f(\tau)$ set equal to 1, the 
 term $-0.65 \tau$ is equivalent to multiplying
$\cpeak$ by the usual  factor $\exp(-\tau)$. 
We use the following formula, which 
was obtained by fitting the output of CMBfast, and is
accurate to a few percent over the interesting range of $\tau$;
\be
f=1- 0.165 \tau/(0.4+\tau)
\,.
\ee

For the running-mass model, we start with  the above estimate for $n=1$,
and    adjust it using \eq{kofell}. Adopting the COBE
normalization mentioned earlier, this adjustment is
\be
\frac{\cpeak}{
\sqrt{\widetilde C\sub{peak}\su{(n=1)}}
} = \frac{\delta_H(k(\ell,\Omega_0))}{\delta_H(k\sub{COBE}(\Omega_0))}
\,.
\label{peakpresc}
\ee
In the case of constant $n$, this  prescription
 corresponds to the previous one with $a_n=0.91$, in good agreement
with the output of CMBfast.

\section{Constant spectral index}

\subsection{The observational constraints}
Most models of inflation make $n$ roughly scale-independent, over the
cosmologically interesting range. We  therefore begin by considering 
the case that $n$ is exactly scale-independent.
 The resulting bound
on $n$ is shown in  Figure 1. In the left-hand panel we make the traditional
assumption that reionization occurs at some fixed redshift $z\sub R$.
In the right-hand panel we make the more reasonable assumption, that it 
occurs when some fixed fraction $f$ of the matter collapses, in a
reasonable range $10^{-4.5}<f<1$. The bounds in the latter case
are relatively insensitive to $f$, because the corresponding range
of $z\sub R$ is narrower; everywhere on the displayed curves, $z\sub R$
is within (usually well within)  the range  $8$ to $36$. 
Details of the fit for $z\sub R=20$ are given in 
Table 1. Practically the same fit is obtained if instead we fix $f$ at
$10^{-1.9}$.

The least-squares fits were  performed with the CERN minuit package,
and the quoted error bars invokes the usual parabolic approximation
(i.e., it they are  the diagonal elements of the error matrix). The
exact error bars given by the same package agree to better than
10\%. For  $z\sub R$, our results  are similar to those  obtained in 
\cite{dick},
but more precise because of improvements in our knowledge of the cosmological
parameters; they  are also similar to those obtained in
 \cite{ruth}, if we take the errors to be the ones given by the error
matrix. (We do not know  why the exact 
 error bars in \cite{ruth} are about three times bigger, in conflict with
both our work and that of \cite{dick}.)
 
After we completed this work, 
 the BOOMeranG \cite{boom} and MAXIMA \cite{maxima} measurements of
the cmb anisotropy appeared, both of which  extend to the second
 acoustic peak. Fits to  these data \cite{boomfit,maximafit}
 seem to again give a similar
constraint on $n$, but the values for $\Omega\sub b$, $\Omega\sub c$
and $h$ outside our adopted 2-$\sigma$ range. 
At the time of writing, the new cmb data  have not been included in a fit
of the type that we are performing (i.e., with 
with strong prior requirements on the cosmological parameters,
as well as on the 
small-scale data  $\widetilde\sigma_8$ and $\widetilde \Gamma$).

\begin{figure}
\centering
\leavevmode\epsfysize=6.5cm \epsfbox{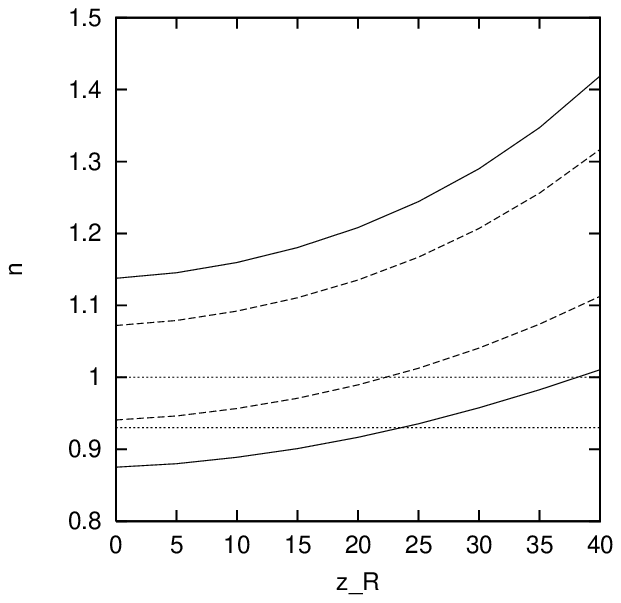} 
\epsfysize=6.5cm \epsfbox{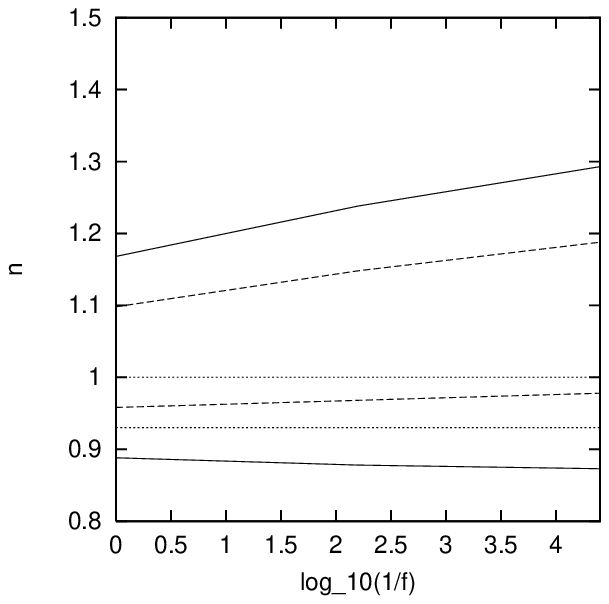}\\
\caption{The nominal 1- and 2-$\sigma$ bounds on $n$.
In the left-hand panel, the reionization redshift
 $z\sub R$ is fixed. In the right-hand panel, reionization is assumed to 
occur when a fixed  fraction $f$ of matter collapses 
(corresponding reionization redshift, not shown, is roughly in the range $10$
to $35$).
A result $n>1$ would rule out  most known models of inflation,
 a result $n>.93$ would rule out   'new' inflation with a cubic potential;
these cases are indicated by horizontal lines. 
}
\end{figure}

\subsection{Models of inflation giving $n<1$}
Although the quality and quantity of data are insufficient for a 
proper statistical analysis, these bounds on $n$ are very striking
when compared with theoretical expectations.
These  expectations  \cite{treview,abook} are summarized\footnote
{This Table  excludes the running-mass models to be discussed later,
and a recently-proposed model \cite{ewan00} giving $(n-1)/2= -2/N$.
It also excludes the ad hoc 'chaotic inflation' potentials
$V\propto \phi^p$, which give  $n-1=-(2+p)/(2N)$ with a 
significant gravitational
contribution to the cmb anisotropy.}
 in Tables 2
and 3, and we now discuss them 
beginning with the usual case $n<1$ (red spectrum). Details of the models
and references are given in \cite{treview}.
 
The simplest prediction is for a potential of the form\footnote
{In this expression and in \eqs{pot}{runpot}, the 
 remaining terms  are supposed to be negligible, and 
$V_0$ is supposed to dominate,
while cosmological scales leave the horizon.}
\be
V=V_0-\frac12 m^2\phi^2 + \cdots
\,,
\label{invq}
\ee
leading to $n-1=-2\mpl^2 m^2/V_0$.
This is the form that one expects if $\phi$ is a string modulus
(Modular Inflation), or
a pseudo-Goldstone boson (Natural Inflation), or  the radial part
of a massive field spontaneously breaking a symmetry (Topological
Inflation). The vacuum expectation value (vev) of $\phi$ in these
models is expected to be of order $\mpl$ or less, while the potential
\eq{invq} gives 
$\vev{\phi}\sim (1-n)^{-1/2}\mpl$.
Therefore, the 
 present bound $n\gsim 0.9$ is already beginning to disfavor these models.
The potential \eq{invq} may however  give $n$ very close to 1 if the potential
steepens after cosmological scales leave the horizon, for instance in an
inverted 
hybrid inflation model.

\begin{table}
\begin{center}
\begin{tabular}{|c|c|ccc|ccc|}
& $n$ & $\omb h^2$ & $\Omega_0$ & $h$ 
&$\widetilde \Gamma$ & $\widetilde \sigma_8$ & $\cpeak$ \\[4pt]
data & --- & $0.019$ & $0.35$ & $0.65$ &
 $0.23$ & $0.56$ & $80\muK$ \\[4pt]
error & --- & 0.002 & 0.075 & 0.075  & 0.035 & 0.059 & $10\muK$
 \\[4pt]
fit & $1.064$ & $0.019$ & $0.34$ & $0.63$
 & 0.19 & 0.59 & $77\muK$ \\[4pt]
error & 0.077 & 0.002 & 0.06 & 0.06  & --- & --- & --- \\[4pt]
$\chi^2$ & --- & $9\times10^{-5}$ & $3\times 10^{-2}$ & 0.1 &
 $1.3$ & $0.2$ & $0.1$ \\[4pt]
\end{tabular}
\label{table1}
\caption{Fit  of the $\Lambda$CDM model to presently available data,
with $z\sub R=20$.
The spectral index $n$ is a parameter of the model, and so are
the next three quantities. Every quantity except $n$ is 
a data point, with the value and uncertainty listed in
the first two rows. The result of the  least-squares fit is given in the
lines three to five.  All uncertainties are at the nominal 1-$\sigma$
level. The total $\chi^2$ is 1.8 for two degrees of freedom.}
\end{center}
\end{table}

\begin{table}
\begin{center}
\label{table2}
\caption{Predictions for the spectral index $n(k)$.
Wavenumber $k$ is related to number of $e$-folds $N$
by $d\ln k=-dN$. 
Constants   $q$ and $Q$ are positive, 
and   $p$ can have either sign.}
\begin{tabular}{|lll|}
Comments 
& $V(\phi)/V_0$ & $\frac12 (n-1)$ \\[4pt] \hline
Mass term & $1\pm\frac12 \frac{m^2}{V_0} \phi^2$ 
& $\pm \mpl^2 m^2/V_0$ 
\\[4pt]
$p$ integer $\leq -1$ or $\geq 3$ 
 & $1+|c|\phi^p$ &
$\frac{p-1}{p-2}\frac1{N\sub{max}-N}$  \\[4pt]
Spont. broken susy &
$1+|c| \ln\frac\phi Q$ & $-\frac1{2N} $  \\[4pt]
Various models & $1- e^{-q\phi}$ & $-\frac1 N$  \\[4pt]
$p>2$ or $-\infty<p<1$ & $1-|c|\phi^p$ &
$-\(\frac{p-1}{p-2} \) \frac1 N$   
\end{tabular}
\end{center}
\end{table}

\begin{table}
\begin{center}
\label{table3}
\caption{ Predictions for the spectral index $n$,
 for some potentials of the form
$V_0(1 +c \phi^p)$ with {\em negative} $c$.
The case $p\to 0$ corresponds to the potential $V_0(1+c\ln\frac{\phi}Q)$,
 and the case
$p\to -\infty$ corresponds to $V_0(1-e^{-q\phi})$.
The parameter $\ncobe<60$ depends on the cosmology after inflation.}
\begin{tabular}{|cll|}
$p$ & $n$ &   \\ 
 & $\ncobe=50$ & $\ncobe=20$  \\ \hline
$p\to 0$ & $0.98$ & $0.95$ \\
$p=-2$ & $0.97$ & $0.93$ \\
$p\to \pm \infty$ & $0.96$ & $0.90$ \\
$p=4$ & $0.94$ & $ 0.85$ \\
$p=3$ & $0.92$ & $ 0.80$  \\ 
\end{tabular}
\end{center}
\end{table}

Of the  remaining models of Table 2, those giving  a red spectrum
 involve a potential basically of the form
\be
V=V_0\( 1+c \phi^p + \cdots \)
\,,
\label{pot}
\ee
with $c$ {\em negative} and $p$ {\em not} in the range $1\leq p \leq 2$.
( 'New' inflation corresponds
to $p$ an integer $\geq 3$, while mutated hybrid inflation models
 account for the
rest of the range.
The logarithmic and exponential potentials in Table 2
 may be regarded as the limits
respectively $p\to 0$ and $p\to - \infty$.)
 With this form, the prediction is
\be
n-1=-\(\frac{p-1}{p-2} \) \frac 2 N
\,.
\label{pred}
\ee
For the moment, we ignore the mild scale-dependence and set
$N=\ncobe$.

 Depending on the history of the Universe, 
\be
N\sub{COBE} \simeq 60 - \ln(10^{16}\GeV/V^{1/4}) - \frac13\ln(V^{1/4}/T\sub{
reh})
-%\Delta N
N_0
\, .
\label{Ncobe}
\ee
In this expression, $T\sub{reh}$ is the  reheat temperature, 
while the final contribution  $-N_0$ (negative in all reasonable cosmologies)
 encodes our ignorance about
what happens between the end of inflation and nucleosynthesis.
Let us pause to discuss this ignorance.
In the present  context, we are  defining  $T\sub{reh}$
 as the temperature when the Universe {\em first}
becomes radiation dominated after inflation. In the conventional
cosmology, radiation domination persists until the present matter
dominated era begins, long after nucleosynthesis. If this is the case,
and if also slow-roll inflation gives way promptly to matter domination
as is the case in most models,
then $N_0=0$.\footnote{In some inflation  models, slow-roll is followed 
by an extended era
of fast-roll giving $N_0$ of order a few; for simplicity we ignore
that possibility in the present discussion.}
 In this conventional  case, $N\sub{COBE}$ is largely  determined by 
$V_0^{1/4}$, and hence by the model of inflation. It is  certainly in the range
$32$ to $60$ (lower limit corresponding to $V_0^{1/4}=100\GeV$) and
much more likely in the range $40$ to $60$ (lower limit corresponding to
$V_0^{1/4}\sim 10^{10}\GeV$ and $T\sub{reh}\sim 100\GeV$).

However, the  conventional cosmology need not be correct. In particular,
the initial radiation-dominated era may give way to matter domination
by a late-decaying particle, and most crucially there may be an era
of thermal inflation \cite{thermal} during the  transition. 
This unconventional cosmology, with its huge entropy dilution after
inflation, is  indeed  demanded in many inflation models, if
 gravitinos created from the vacuum fluctuation \cite{gravitino}
persists to late times \cite{latetime}.
Even one bout of thermal inflation will give $N_0\sim 10$ and additional
bout(s) cannot be ruled out. Thus, from the theoretical viewpoint, 
$N\sub{COBE}$ can be anywhere in the range $0$ to $60$.

Let us discuss the prediction \eq{pred}, excluding for simplicity the
ranges $0<p<1$ and $2<p<3$ (recall that the  straightforward
'new' inflation models make $p$ an integer $\geq3$).
Taking the maximum value  $\ncobe\simeq 60$, we learn that
 $n<0.93$ for $p=3$ (the lowest prediction), and $n<0.95$ for $p=4$.
Looking at the right-hand panel of Figure 1, we see  that at 
nominal 1-$\sigma$ level,
the former case is ruled out, though it is still allowed
at the 2-$\sigma$ level.
Stronger results hold in the if $\ncobe < 60$. Looking at things another
way,  a lower bound on $n$ gives a lower bound on $\ncobe$,
\be
\ncobe >\frac{p-1}{p-2}\,\frac2{1-n} \\
\,.
\ee
Even with present data, the  2-$\sigma$ result $n\gsim .9$ gives
$\ncobe\gsim 40$ for $p=3$, and $\ncobe\gsim20$ for $p\gg 3$.

The scale dependence given by \eq{pred} is
\be
\frac{\diff n}{\diff \ln k} = -\frac12\(\frac{p-2}{p-1}\) \(n-1\)^2
<0
\label{scaledep1}
\,.
\ee
Over  the cosmological
range of scales $\ln(k/k\sub{COBE})$ is at most a few, and in particular
 $\ln(8^{-1} h\Mpc^{-1}/k\sub{COBE})\simeq 4$, corresponding to
\be
\Delta n\equiv
n_8-n\sub{COBE} = -.02 \(\frac{p-2}{p-1} \) \(\frac{n-1}{0.1}\)^2
<0
\,.
\label{scaledep2}
\ee
Taking    $n = 0.9$ to saturate the present bound,
this gives 
 $|\Delta n|<0.02$
with $p\geq 3$, and 
$|\Delta n|<0.04$ with $p\leq 0$.
Even in the latter case, the change in $n$ is hardly
 significant with present data.

\subsection{Models giving $n>1$}

Known models 
 giving $n>1$ (blue spectrum) are all of the 
hybrid inflation type. The simplest case is $V=V_0+\frac12m^2\phi^2$;
it  gives the scale-independent prediction $n-1=2\mpl^2m^2/V_0$,
which may be either close to 1 or well above 1.

The other cases
involve a potential of the form $
V=V_0\( 1+c\phi^p \)$ with  {\em positive} $c$, and $p$ an integer
$\geq 3$ or $\leq -1$.
 There is  a maximum (early-time) value
for $N$, and the prediction
\be
n-1 = \frac{p-1}{p-2} \frac1{N\sub{max} - N}
\,.
\ee
Barring the fine-tuning $N\sub{COBE}\simeq N\sub{max}$,
this gives $n-1\ll 0.04$, which is compatible with the 
observational bound. 
The scale-dependence of $n$ in these models is still given by 
\eqs{scaledep1}{scaledep2}; it  may be observationally significant
only in the fine-tuned case $N\sub{COBE}\simeq N\sub{max}$,
which we have not investigated. 

\section{The running mass models}

\subsection{The potential}
We have also done fits with the  
 scale-dependent spectral index,
 predicted in inflation models with a running inflaton mass
 \cite{st97,st97bis,clr98,cl98,c98,rs}. In these models,
based on softly broken supersymmetry,
one-loop corrections to the tree-level potential are taken into
account, by evaluating the inflaton mass-squared
 $m^2(\ln (Q))$ at the renormalization 
scale $Q\simeq \phi$,\footnote
{The choice $Q\simeq \phi$  is to be made in the regime
where $\phi$ is bigger than the  relevant masses. When $Q$ falls below 
the relevant masses, 
 $m^2(Q)$ becomes practically  scale-independent (the mass 'stops running').
We have a running mass model  if inflation takes place in the former
regime, which happens in some interesting cases \cite{cl98,c98},
including that of a  gauge coupling.}
\be
V=V_0 + {1\over 2} m^2(\ln(Q)) \phi^2 + \cdots
\,.
\label{runpot}
\ee

Over any  small  range of $\phi$, 
it is a good approximation to  take the running mass to be  a 
linear function of $\ln\phi$.
This is equivalent to choosing the renormalization scale to be 
within the range, and then adding the loop correction explicitly,
\be
V=V_0 +\frac12m^2(\ln Q)\phi^2 -\frac 12 c(\ln Q) \frac{V_0}{\mpl^2}
\phi^2 \ln(\phi/Q)
\,.
\label{vlin1}
\ee
The dimensionless quantity  $c$ specifies the strength of the coupling.
Let us discuss its likely magnitude, taking for definiteness
 $Q=\phi\sub{COBE}$.

It has been shown \cite{cl98} that the linear approximation
is very good over the range of $\phi$ corresponding to horizon exit
for scales between $k\sub{COBE}$ and $8h^{-1}\Mpc$. We shall want
to estimate the reionization epoch, which involves a
 scale of order $k\sub{reion}^{-1}\sim
10^{-2}\Mpc$ (enclosing the relevant mass of order
$10^6\msun$). Since only a crude estimate of the reionization
epoch is needed, we shall assume that the linear approximation is
adequate down to this `reionization scale'. In other words,
we assume that it is adequate for $\phi$ between $\phi\sub{COBE}$
and $\phi\sub{reion}$, the subscripts denoting the value of $\phi$
when the relevant scale leaves the horizon.
Within this range, we it is  convenient to write \eq{vlin}
in the form \cite{cl98}
\be
V=V_0-\frac12 \frac{V_0}{\mpl^2} c\phi^2\( \ln\frac{\phi}{\phi_*}
-\frac12 \) 
\,,
\label{vlin}
\ee
 so that
\be
V'=- \frac{V_0}{\mpl^2} c\phi  \ln\frac{\phi}{\phi_*}
\,.
\ee
In these  expressions, the constants $c$ and $\phi_*$ both depend on the 
renormalization scale
$Q$, which can be chosen 
 anywhere in the range corresponding to cosmological
scales (say $Q=\phi\sub{COBE}$).
 The dimensionful constant $\phi_*$ is related to 
 the  mass-squared by
\be
\ln(\phi_*/Q) = {m^2(Q)\mpl^2\over c(Q) V_0}  -\frac12
\,.
\label{mofphi}
\ee
Note that the limit of no running, $c \rightarrow 0$, corresponds to
finite $ c |\ln(\phi/\phi_*)| $, so that \eq{vlin} in that
limit gives back \eq{runpot} with a constant mass.

In general, the point $\phi=\phi_*$  may  be far
 outside the  regime
where the linear approximation \eq{vlin} applies.
 However, in simple models the cosmological 
regime is sufficiently close to that point 
that the linear approximation 
is approximately valid there. 
In that case, we can trust the 
 \eq{vlin}  and its derivatives for $\phi=\phi_*$; 
since $V'$ vanishes at that point,
 there are four
clearly distinct models of inflation as shown in
Figure \ref{models}. The labeling  (i), (ii), (iii) and (iv)
is the one introduced 
in \cite{cl98}.
In Models (i) and (ii), 
 $c$ is positive and the potential
has  a maximum near $\phi_*$,
while in Models (iii) and (iv),
$c$ is negative and there is a minimum.
In Models (i) and   (iv),
 $\phi$ moves towards the origin, while  in Models
(ii) and   (iii) the opposite is true.
Even if \eq{vlin} is not valid near $\phi=\phi_*$, 
this fourfold classification of models,  according to the sign of $c$ and
the direction of motion of $\phi$, is still useful.

Let us discuss the likely magnitude of $c$, assuming that 
 a single coupling dominates the loop correction.
 The value of $c$ is conveniently obtained
 from the well-known
RGE for $\diff m^2/\diff (\ln Q)$.
 If a gauge coupling dominates one finds
\cite{st97bis}
\be
\frac{V_0 c}{\mpl^2} =  \frac{2 C}\pi \alpha \widetilde m^2
\,.
\label{c-def}
\ee
Here, $C$ is a positive  group-theoretic number of order 1, 
 $\alpha$ is the gauge coupling, and 
 $\widetilde m$ is the gaugino mass.
 We see that
 {\em if the loop correction comes from a single gauge
coupling, $c$ is positive}, corresponding to Model (i) or Model
(ii). If a Yukawa coupling dominates, one finds \cite{c98} (for
negligible supersymmetry breaking trilinear coupling)
\be
\frac{V_0 c}{\mpl^2} = - {D\over 16\pi^2} |\lambda |^2 m^2_{loop}
\,,
\ee
where $D$ is  a positive constant counting the number of 
scalar particles interacting with the inflaton, $m^2\sub{loop}$ 
is their common susy breaking mass-squared, and 
$\lambda$ is their common  Yukawa coupling.
 In this case, $c$ can be of either
sign. 

To complete our estimate of $c$, we 
 need  the gaugino or scalar mass. 
  The  traditional
hypothesis is that soft supersymmetry breaking is gravity-mediated,
and in the context of inflation this means that the scale
$M\sub S$  of supersymmetry
breaking will be  roughly $V_0^{1/4}$. (As usual we are defining 
$M\sub S\equiv \sqrt F$, where $F$ is the auxiliary field responsible
for spontaneous supersymmetry breaking  in the hidden sector.
 We also assume that there is no accurate cancelation
in the formula $V=|F|^2-3\mpl^2m_{3/2}^2$, which is the case in most
supersymmetric inflation models \cite{treview}.)
With gravity-mediated susy breaking, typical values of the 
 masses are $\widetilde m^2\sim |m\sub{loop}^2|\sim V_0/\mpl^2$,
which makes $|c|$ of order of the coupling strength
  $\alpha$ or  $|\lambda|^2$. At least in  the case of a gauge coupling, one
then expects
\be
|c|\sim 10^{-1}\mbox{ to }10^{-2}
\,.
\ee
In special versions of gravity-mediated susy breaking,
 the masses could be much smaller, leading to $|c|\ll 1$. In that case,
the mass would hardly run, and the spectral index would be practically
scale-independent. With gauge-mediated
susy breaking, the masses  could be much bigger; this
 would not
lead to a model of inflation
(unless the coupling is suppressed) because
it would not satisfy the  slow-roll requirement
$|c|\lsim 1$. 

\subsection{The spectrum and the  spectral index}

Using 
\eq{Nofv}  we find
\bea
s e^{c\Delta N(k)} &=& c \ln(\phi_*/\phi)\label{sigma}\\
\Delta N(k)&\equiv& N\sub{COBE} - N(k)
\equiv \ln(k/k\sub{COBE})
\,,
\eea
where 
$s$ is an integration constant.\footnote
{In an earlier paper \cite{cl98}
we used $\sigma\equiv s e^{cN\sub{COBE}}$, 
but $s$ is more convenient.}
 \eq{nofvapprox} then gives
\be
{n(k)-1\over 2} = 
s e^{c\Delta N(k)} - c \label{runpred}
\,.
\ee
Some lines of  fixed $n\sub{COBE}$ in the plane $s$ versus $c$
are shown in the left-hand panel of Figure \ref{s-c-f1}.
In order to evaluate \eq{peakpresc}, we also need the variation of
$\delta_H$ which comes from integrating this expression,
\be
\frac{\delta_H(k)}{\delta_H(k\sub{COBE})}
=\exp\[ \frac sc \(e^{c\Delta N}-1\)-c\Delta N \]
\,.
\ee

We are mostly interested in cosmological scales between $k\sub{COBE}$
and $k_8$, corresponding to $0\lsim \Delta N\lsim 4$.
In this range the scale-dependence of $n$ is approximately linear
(taking $|c|\lsim 1$)
and the variation $\Delta n\equiv n_8-n\sub{COBE}$ is given approximately
by
\be
\Delta n \simeq 4
\frac{\diff n}{\diff \ln k}\simeq 8sc
\,.
\label{scaledep3}
\ee
In contrast with the  prediction 
 \eqs{scaledep1}{scaledep2} of the earlier models we considered,
$\Delta n$ is positive. Also in contrast with those models,
it is  not tied to the magnitude of $|n-1|$, and (as we shall see)
 may be significant even with present data, for physically reasonable
values of the parameters.
        In the right-hand panel of 
 Figure \ref{s-c-f1},
we show  the branches
of the hyperbola $8sc=\Delta n$, for the reference
value
$\Delta n=0.04$.
 Within
the hyperbola, the scale-dependence of $n$ is probably 
too small to be significant
with present data.

\begin{figure}
\centering
\leavevmode\epsfysize=6.5cm \epsfbox{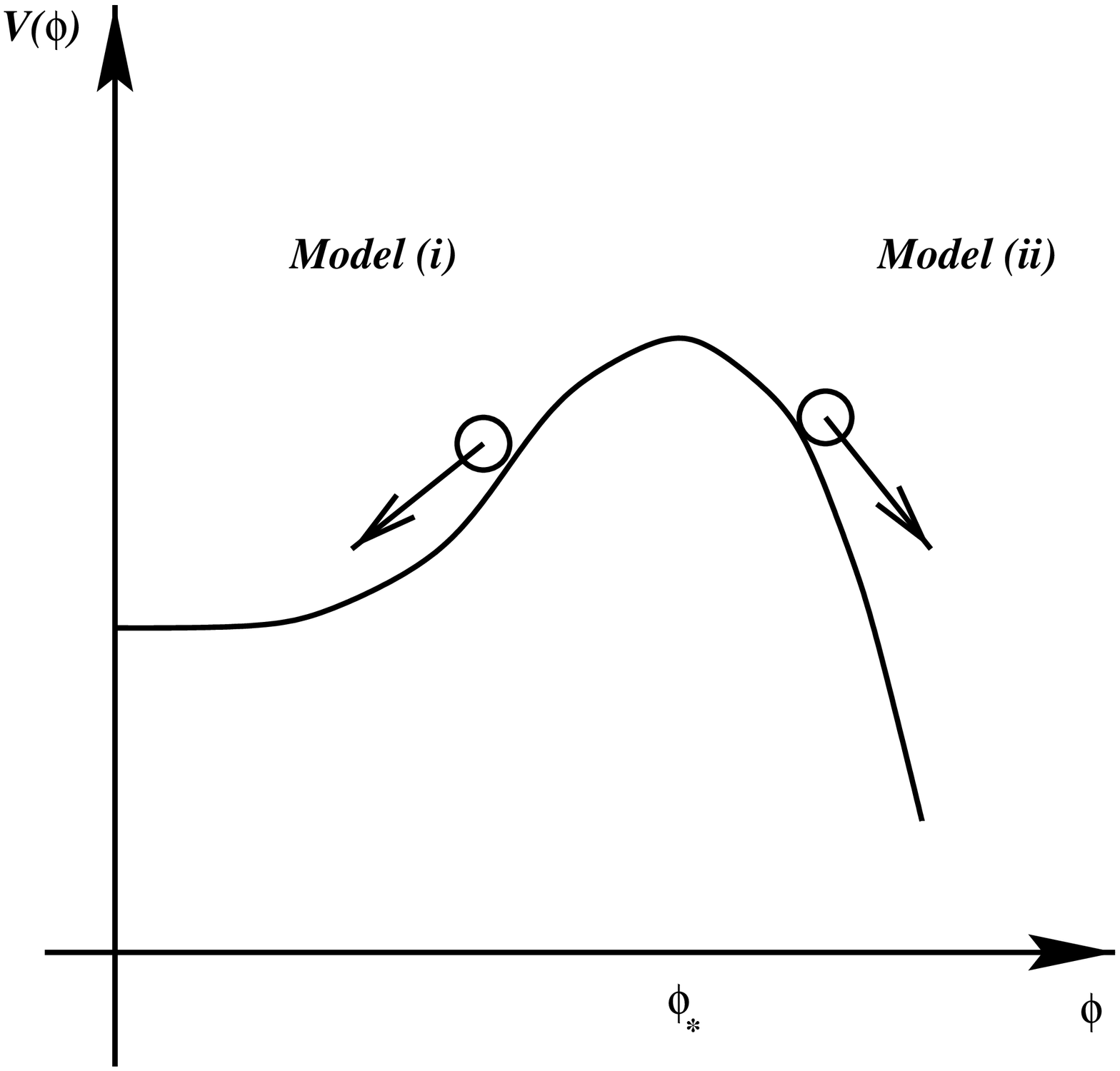}
\epsfysize=6.5cm \epsfbox{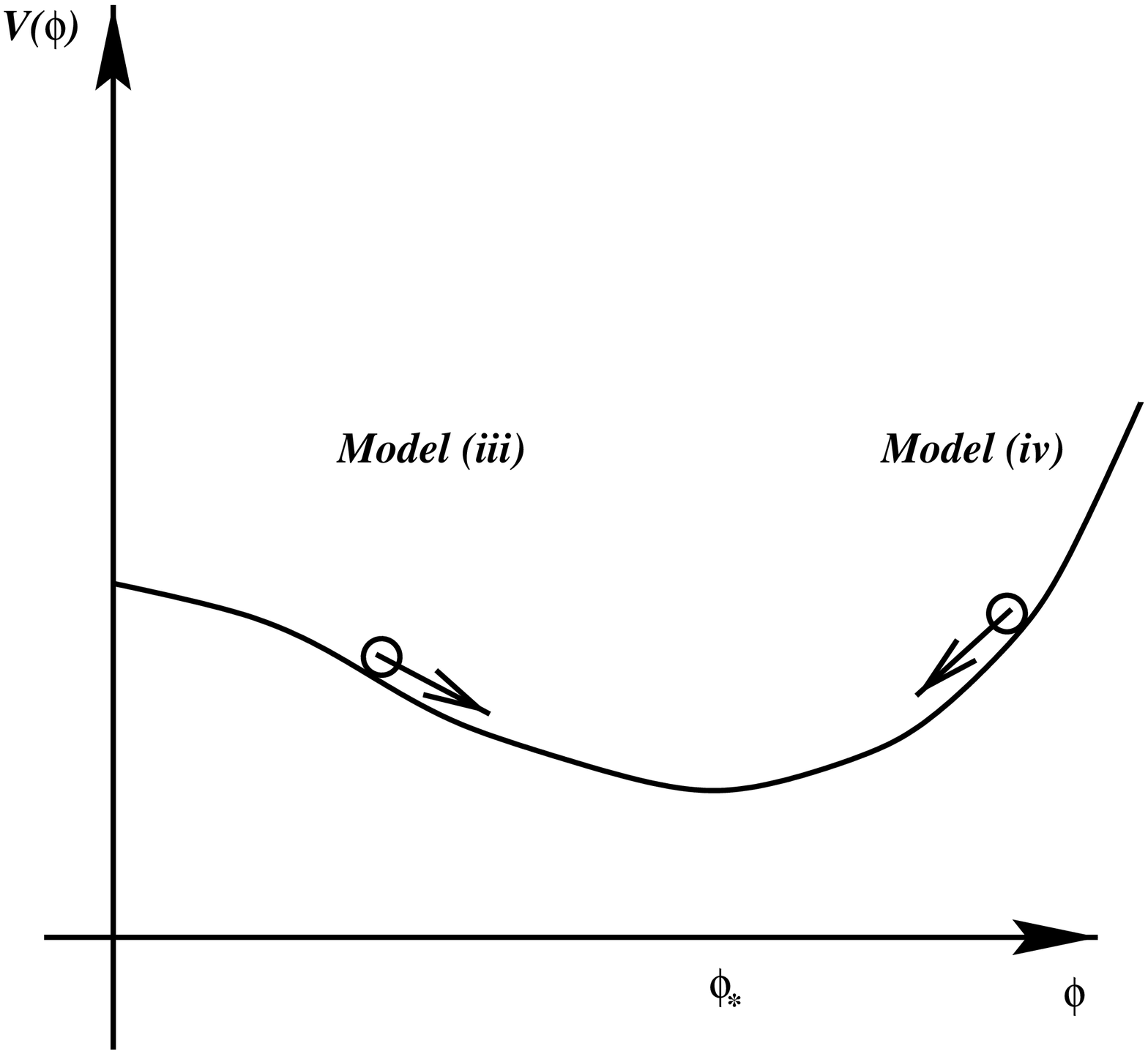}\\
\caption[sc-fig1]{Sketches of the potential for the different models
in the case an extremum exists: the right panel shows the inflaton 
behavior for Models (i) and (ii), while the left panel shows
Models (iii) and (iv).}
\label{models}
\end{figure}

The spectral index  \eq{runpred} depends on the coupling $c$,
which we already discussed, and the integration constant $s$.
To satisfy the  slow-roll conditions $\mpl^2|V''/V|\ll 1$ and
$\mpl^2(V'/V)^2\ll1 $,
 both $c$ and $s$ 
must be at most of  order 1 in magnitude.
Significant additional 
 constraints on $s$ follow, if we make the reasonable
assumptions that
 the mass continues to run to the end of 
slow-roll inflation, and that 
 the   linear approximation
 remains  {\em roughly} valid.  Indeed, setting 
$\Delta N=N\sub{COBE}$, \eq{sigma} becomes
$
s=e^{-c\ncobe}c\ln(\phi_*/\phi\sub{end})$.
Discounting the possibility that 
 the end of inflation is very fine-tuned, to occur close to the
maximum or minimum of the potential, this 
gives a  lower bound 
\be
|s|\gsim e^{-c\ncobe}|c|
\,.
\label{sb1}
\ee
In the case of positive $c$ (Models (i) and (ii)),
 we also obtain a significant upper bound
by 
setting $\Delta N=\ncobe$ in \eq{runpred}, and remembering
that slow-roll requires
 $|n-1|\lsim 1$;
\be
|s|\lsim e^{-cN\sub{COBE}}\hspace{2em}(c>1)
\,.
\label{sb2}
\ee
In the simplest case, that 
slow-roll 
inflation 
  ends when $n-1$ actually becomes of order 1, this bound becomes
an actual estimate,
 $|s|\sim e^{-cN\sub{COBE}}$.

 In the case of 
Models (i) and (iv), 
the mass may cease to run before the end of slow-roll inflation
(but after cosmological scales leave the horizon, 
 or the running mass model
would not apply) at some point $N\sub{run}$.
 In this somewhat  fine-tuned situation,
$\ncobe$ in the above estimates should
be replaced $\ncobe-N\sub{run}$, which may be much less than $\ncobe$.
In the case of Model (iv), this leads to a weaker 
 lower bound
\be
s\gsim  |c|\hspace{3em}(c<0)
\,.
\label{sb3}
\ee
In the case of Model (i) it leads to a weaker
upper bound
\be
s\lsim 1\hspace{3em}(c>0)
\,.
\label{sb4}
\ee
In the left-hand panel of Figure \ref{s-c-f1},
we show   the bounds relevant to the 
choice of parameter's ranges, i.e. the 
lower bound \eq{sb1}, the upper bound \eq{sb2}
and the weak lower bound \eq{sb3}.

\subsection{The magnitude of the spectrum}
Although it is not directly relevant for our investigation of the spectral
index, we should mention the constraint on the running mass model that comes
from the observed magnitude $\calpr^{1/2}\simeq 10^{-5}$ of the spectrum.
{}From \eq{delh}, 
\be
\frac4{25}\calpr=
 \frac{V_0}{\phi_*^2\mpl^2}\exp\(\frac
{2s}c \)\frac1{|s|^2} 
\,.
\label{cobenorm}
\ee
This prediction 
 involves $V_0$ and $\phi_*$, in addition to the parameters $c$
and $s$ that determine the spectral index.

The simplest thing is to again assume gravity-mediated susy breaking,
with the ultra-violet cutoff at the traditional scale around $\mpl$,
and the same supersymmetry breaking scale
 during inflation as in the true vacuum so that
 $V_0^{1/4}\sim 10^{10}\GeV$. 
In this scenario, one expects
 $|m^2(Q)|\sim V_0/\mpl^2$ at $Q\sim \mpl$.
As Stewart pointed out in the first paper
on the subject, with this very traditional set of assumptions,
\eq{cobenorm} can give  the correct COBE normalization, with 
 $|c|$ in the physically favored range $10^{-1}$ to $10^{-2}$.\footnote
{At the crudest level, one can verify this  using
the  linear approximation \eq{vlin}
 all the way up to the $\phi\sim\mpl$,  corresponding to 
$\ln(\mpl/\phi_*)\sim 1/c\sim 10$ to $100$. Proper calculations
\cite{st97bis,clr98,cl98}
using the RGE's lead to the same conclusion.}

It is remarkable that 
 the most traditional set of assumptions can give a model with the 
correct COBE normalization, and, as we shall see, with  
a viable spectral index.
If one relaxes these assumptions, 
  there is much more freedom in choosing
 $V_0$ and $\phi_*$. Such
  freedom
may be very welcome, in coping with the difficulty of implementing
 inflation in the context of  large extra dimensions \cite{large}.

\subsection{Observational constraints on the running mass models}

\begin{figure}[t]
\centering
\leavevmode
\epsfysize=7.5cm \epsfbox{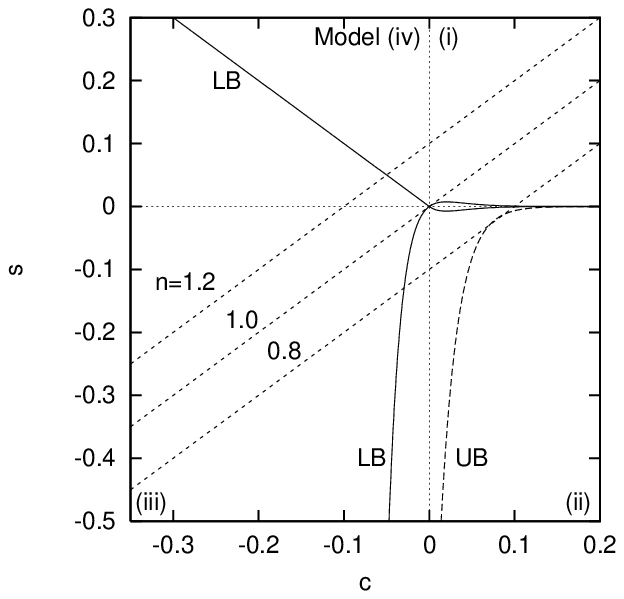}
\epsfysize=7.5cm \epsfbox{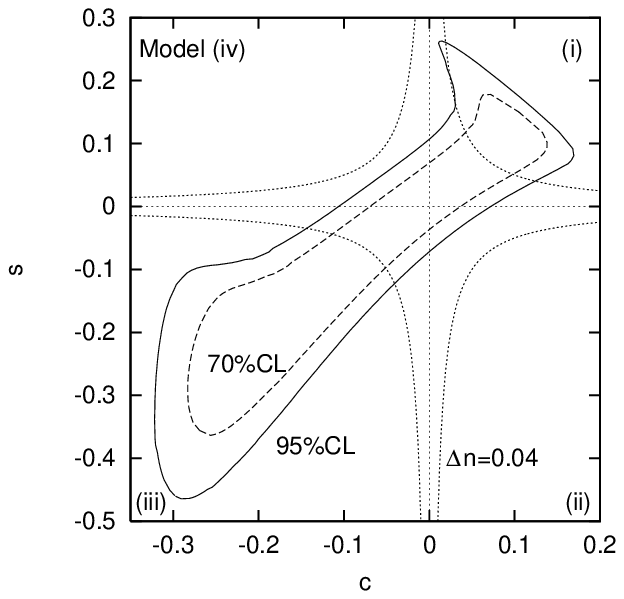}
\caption[sc-fig2]{
The  parameter space for the running mass model.
In the  left-hand panel we show the straight lines 
corresponding to $n\sub{COBE}=1.2$, $1.0$ and $0.8$. 
Also shown in the left-hand panel are the lower  bound \eq{sb1},
the upper bound \eq{sb2}, and (diagonal line in upper right quadrant)
the weak lower bound \eq{sb3}. (The weak upper bound \eq{sb4} is off the 
scale.) As explained in the text, these curves 
define the theoretically reasonable region of the parameter space.
In the right-hand panel, we show the region allowed by observation,
in  the case that reionization occurs when  $f\simeq 1$.
Note that the allowed region is parallel to the fixed $n_{COBE}$
lines around $n_{COBE}\simeq 1$, as one would expect.
To show the scale-dependence of the prediction for $n$, we also show in 
this panel the branches
of the hyperbola $8sc=\Delta n\equiv n_8-n\sub{COBE}$, for the reference
value $\Delta n=0.04$.}
\label{s-c-f1}
\end{figure}

Extremizing with respect to all other parameters, we have calculated
$\chi^2$ in the $s$ vs. $c$ plane and obtained contour levels
for $\chi^2$ equal to the minimum value plus
$2.41$ and $5.99$ respectively, 
corresponding nominally to the 70\% and 95\% confidence level in 
two variables. (The
$\chi^2$ function presents actually two nearly degenerate minima in the 
allowed region, one in the positive and one in the negative quadrants 
(Models (i) and (iii)), separated by a very low barrier, but we  assume
that the usual quadratic estimate of the probability content is not very 
far from the true value.)

The allowed region is shown in the right-hand panel of 
Figure \ref{s-c-f1}, for
 the case that 
 reionization occurs when  $f\simeq 1$.
For $c=0$ or $s = 0$ the constant $n$ result is recovered with
$n-1= -2 c$ or $ 2s$; our plots give in this case a slightly larger 
allowed interval with respect to the two sigma value in the previous
section, due to the mismatch between the statistical one variable and
two variables 95\% CL contours.
 This allowed region is not too different
from  the one that we estimated
earlier \cite{cl98}, by imposing the crude requirement
$|n-1| < 0.2 $ at both the COBE scale and the low scale 
corresponding to $N_{COBE} - 10$.
(Note that in the earlier work we used the less convenient
variable  $\sigma\equiv s\exp(c\ncobe)$, instead of $s$.)
%We can assume that the freedom in changing the cosmological 
%parameters is in some way compensating the more accurate 
%prescription we are using here.

The allowed region for Models (ii) 
 and (iv) lies inside the
hyperbola corresponding to $\Delta n=.04$,
which means that their scale-dependence is  hardly significant
 at the level of present data.
In contrast, the allowed region for Models (i) and (iii)
 extends to $\Delta n\geq 0.2$, representing an extremely significant
scale-dependence even with present data.
To demonstrate this, we show in Figures \ref{n8ncobe} and 
\ref{n8ncobe3} the allowed regions for Models (i) and (iii)
in the $n_8$ versus $n\sub{COBE}$ plane. 
In the case of  Model (iii), 
the theoretical bounds on the
parameters restrict the parameter space to a small corner of
the allowed region, within which $n$ has 
 negligible variation.
 In contrast, there is no significant  theoretical restriction on
the parameters in the case of Model (i), and $n$ has significant variation
in a physically reasonable regime of parameter space.
In both cases, 
 a lower value of the fraction of 
collapsed matter $f$ just reduces the allowed region
at large n, without affecting significantly  the allowed scale-dependence
of n. 

 In the case of Model (i), a further observational constraint comes
from the requirement that the
 density perturbation on scales leaving the horizon at the {\em end}
of inflation, should be small enough to avoid dangerous black hole
formation. The linear approximation is not adequate
on such small scales, and one should instead evaluate the running
mass using the RGE. The simplest assumption is that the RGE
corresponds to a single gauge coupling, either with or without asymptotic
freedom \cite{clr98}. The black hole constraint has been evaluated for these
cases \cite{lgl}. 
The constraint amounts  more or less to an upper bound on
$n\sub{COBE}$, typically in the range $1.1$ to $1.3$ depending on the
choices of $N\sub{COBE}$ and other parameters. Such a bound
significantly reduces the allowed region of parameter space, 
but still leaves a region where $n$ has a strong variation.

\begin{figure}[t]
\centering
\leavevmode\epsfysize=7.5cm \epsfbox{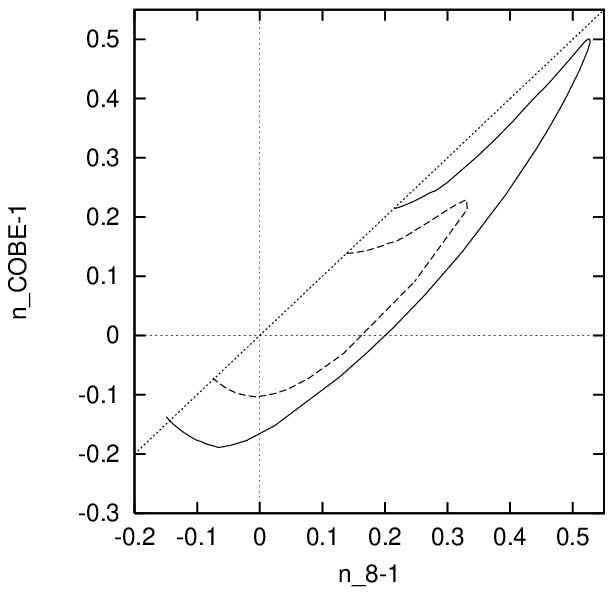}
\epsfysize=7.5cm \epsfbox{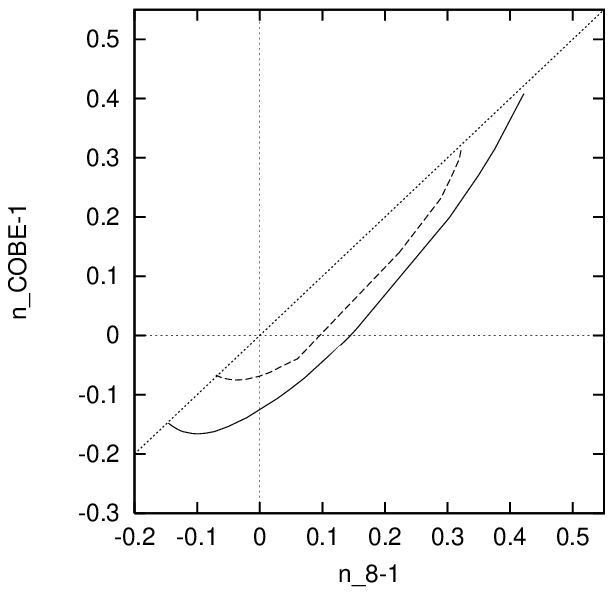}
\caption[n8-ncobe]{Allowed region in the $n_{COBE}-1$ 
vs $n_8-1$ plane at 95\% CL (solid line) and 70\% CL (dashed line)
for positive $s$ and $c$ (Model (i)). The two panels correspond
to different hypotheses about the reionization epoch.
In the right panel, it is assumed that 
reionization occurs when a fraction
 $f = 10^{-2.2}$ of the matter has collapsed into bound structures,
while in the left panel the fraction is taken to be $f\sim 1$. 
}
\label{n8ncobe}
\end{figure}

\begin{figure}
\centering
\leavevmode\epsfysize=7.5cm \epsfbox{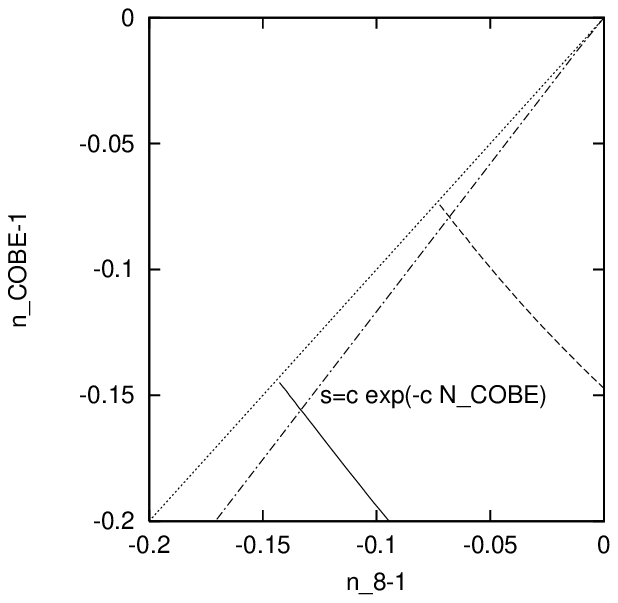}
\epsfysize=7.5cm \epsfbox{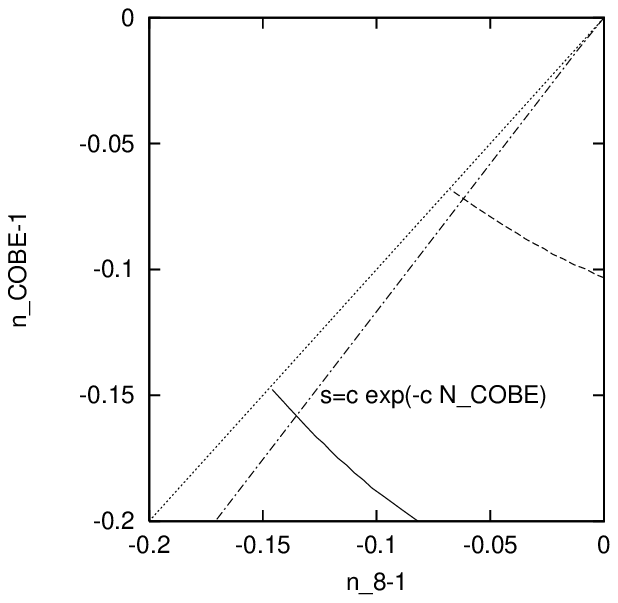}
\caption[n8-ncobe2]{Allowed region in the $n_{COBE}-1$ 
vs $n_8-1$ plane for negative $s$ and $c$ (Model (iii)). 
Again the two panels
correspond to different reionization epoch hypothesis, as in 
Fig.\ref{n8ncobe}. 
 The allowed region is below the dotted line $n_8=n\sub{COBE}$,
and above the solid (dashed) line at 95\% (70\%) confidence level.
These lines  do not depend on the value of 
$N_{COBE}$.
The line $s=c e^{c N_{COBE}}$ is also drawn for 
$N_{COBE}=50$. The theoretically favored
regime   $|s| \geq |c| e^{c N_{COBE}} $ is the
sector between this  line and the $n_8=n_{COBE}$ line.
The region of positive
$n_8-1$ and/or  $n\sub{COBE}-1$ is not shown, since it corresponds to
$|c| \gg |s| e^{-c N_{COBE}}  $.}
\label{n8ncobe3}
\end{figure}

\section{Conclusion}

In the context of the $\Lambda$CDM model, we 
 have evaluated the observational constraint on the  spectral index 
$n(k)$.
This constraint comes from 
 a range of data, including the height of the first peak in the
 cmb anisotropy,  which we take to be
$80\pm 10\mu$K (nominal 1-$\sigma$). 
Reionization is assumed to occur when some fixed fraction $f$ of the 
matter collapses, and the most important results are insensitive
to this fraction in the reasonable range $10^{-4}\lsim f\lsim 1$.

We first considered the  case that $n$ 
 has negligible scale dependence, comparing the observational bound with
the prediction of various models of inflation.
A significant improvement in the 2-$\sigma$ lower bound, which may well occur
with the advent of slightly better measurements of the cmb anisotropy,
will   become a serious discriminator between models of inflation.
Even the present  bound has serious implications if, as is very
possible,  late-time gravitino creation or some other 
phenomenon requires an era of thermal inflation
after the usual inflation.

We also considered  the running mass models of inflation,
 where the spectral
index can have significant scale-dependence. Because of this
scale dependence, it is in this case crucial to fix not the epoch 
of reionization, but the fraction $f$ of matter
that has collapsed at that epoch. We presented results for the choice
$f=1$ (corresponding to $z\sub R\simeq 13$ if the spectral index has
 negligible scale-dependence), and for a perhaps more reasonable
choice  $ f=10^{-2.2}$.
 In the running-mass  models, the
 scale-dependent spectral index  $n(k)$ is given by $n-1=s \exp(c\Delta N)
-c$, where $\Delta N=\ln(k\sub{COBE}/k)$. The parameters in this expression
can be of either sign, leading to four different models of inflation.
Barring fine-tuning, one expects  $s$ to be in the range
$|c|e^{-cN\sub{COBE}}\lsim |s|\lsim e^{-cN\sub{COBE}}$.
The parameter $c$ depends on the nature of the soft supersymmetry
breaking, but in 
 the simplest case of 
 gravity-mediation it becomes a dimensionless coupling strength,
presumably of order 
$ 10^{-1}$ to $10^{-2}$ in magnitude.

Without worrying  about the origin of the parameters
 $c$ and $s$, 
we have investigated the observational constraints  on them.
 In  the  case $c,s>0$ (referred to as Model (i)) we  find that 
$n$ can have a significant  variation on
cosmological scales, with 
 $n-1$ passing  through zero signaling a  minimum of the spectrum
of the primordial curvature perturbation.
In a future paper, we shall exhibit  the 
possible  effect of this scale-dependence
 on the cmb anisotropy, at and above the first peak.

%%%%%%%%%%%%%%%%%%%%%%%%%%%%%%%%%%%%%%%%%%%%%%%%%%%%%%%%%%%%%%%%%%%%%%%%
\section*{Acknowledgments}
We  thank Pedro Ferreira and Andrew Liddle and Martin White
for useful discussions.

%%%%%%%%%%%%%%%%%%%%%%%%%%%%%%%%%%%%%%%%%%%%%%%%%%%%%%%%%%%%%%%%%%%%%%%%

\renewcommand\pl[3]{Phys. Lett. {\bf #1}, #2 (#3)}
\newcommand\np[3]{Nucl. Phys. {\bf #1}, #2 (#3)}
\newcommand\pr[3]{Phys. Rep. {\bf #1}, #2 (#3)}
\renewcommand\prl[3]{Phys. Rev. Lett. {\bf #1}, #2 (#3)}
\renewcommand\prd[3]{Phys. Rev. D{\bf #1}, #2 (#3)}
\newcommand\ptp[3]{Prog. Theor. Phys. {\bf #1}, #2 (#3)}
\newcommand\rpp[3]{Rep. on Prog. in Phys. {\bf #1}, #2,  (#3)}
\newcommand\jhep[2]{JHEP #1 (19#2)}
\newcommand\grg[3]{Gen. Rel. Grav. {\bf #1}, #2,  (#3)}
\newcommand\mnras[3]{MNRAS {\bf #1}, #2,  (#3)}
\newcommand\apjl[3]{Astrophys. J. Lett. MNRAS {\bf #1}, #2,  (#3)}

\end{document}